\documentclass[floatfix,pra,aps,twocolumn,showpacs,preprintnumbers,amsmath,amssymb]{revtex4-1}
\usepackage{float}
\usepackage{amsfonts}
\usepackage{amsmath}
\usepackage{amssymb}
\usepackage{graphicx}
\usepackage{listings}
\usepackage{bm}
\usepackage{bbold}
\usepackage{braket}
\usepackage{todonotes}
\usepackage{hyperref}
\usepackage{url}
\usepackage[utf8]{inputenc}

\begin{document}

\title{The Casimir energy between nanostructured gratings of arbitrary periodic profile}

\author{J. Lussange}
\affiliation{Laboratoire Kastler-Brossel, CNRS, ENS, UPMC, Case 74, F-75252 Paris, France}
\author{R. Gu\'erout}
\affiliation{Laboratoire Kastler-Brossel, CNRS, ENS, UPMC, Case 74, F-75252 Paris, France}
\author{A. Lambrecht}
\affiliation{Laboratoire Kastler-Brossel, CNRS, ENS, UPMC, Case 74, F-75252 Paris, France}

\date{\today}

\begin{abstract}
We study the lateral dependence of the Casimir energy for different
corrugated gratings of arbitrary periodic profile. To this end we
model the profiles as stacks of horizontal rectangular slices
following the profiles' shape and evaluate numerically the Casimir
energy between them for different relative lateral displacements of
the two corrugated plates. We compare our results with predictions
obtained within the proximity force approximation (PFA). At
comparable separation of the corrugated plates and geometric
parameters, we find a strong dependence of the Casimir energy on the
shape of the corrugation profiles.
\end{abstract}

\pacs{03.70.+k, 12.20.Ds, 42.50.Lc}

\maketitle

\section{Introduction}

Two surfaces in quantum vacuum attract each other due to the
fluctuations of the electromagnetic field prevailing in the vacuum
state as predicted by H. Casimir  in 1948 \cite{Casimir1948}. The
Casimir force scales with the inverse of the fourth power of the
surfaces' separation and can take on significant values between
objects in the submicrometer range. Indeed, the dynamics of the
electromagnetic field, including the vacuum field, strongly depends
on its external boundary conditions, and the variation of its
spectral density when the field is confined leads to differences in
the vacuum radiation pressure resulting in a net force.

While the Casimir force is negligible at macroscopic scales it can
become comparable to electric forces at submicron distances. In
microelectromechanical systems (MEMS) it may produce sticking and
adhesion of the system's components \cite{RoukesEPL2001}, but it may
also be used to actuate the system
\cite{ChanScience2001,ChanPRL2001}. Modern measurement techniques
have allowed for accurate determination and measurements of the
Casimir force for different configurations and geometries. Casimir's
original configuration of two flat parallel plates has been
experimentally realized in \cite{BressiPRL2002}. Most of the
measurements have however been performed between flat and spherical
mirrors
\cite{ChanPRL2001,LamoreauxPRL1997,MohideenPRL1998,DeccaPRL2003,LisantiPNAS2005,JourdanEPL2009,TorricelliEPL2011},
and some have employed cylinders \cite{EderthPRA2000,DeccaPRA2011}.

Deviations from the two plates geometry are often accounted for in
calculations by using the proximity force approximation (PFA) which
amounts to adding up local contributions to the Casimir force at
different distances as if those were independent of each other
\cite{DerjaguinKollZ1934}. Exact calculations have shown
considerable deviations from PFA for the plane-sphere geometry
\cite{Canaguier-DurandPRL2009,Canaguier-DurandPRL2010} and when
nanostructured plates are used \cite{LambrechtPRL2008,BaoPRL2010}.
In particular the normal Casimir force is affected by the gratings
geometry, and the change cannot be reliably estimated within PFA as
soon as the different length scales of the problem (corrugation
period $d$, depth $a$ and relative distance $L$) are of the same
order. This is shown by the experiment reported in
\cite{ChanPRL2008} for the interaction of a Au sphere with a Si
grating with deep rectangular trenches. Exact calculations within a
path integral approach of this geometry for perfectly reflecting
plates \cite{BuescherPRA2004} confirm the invalidity of PFA but lead
to too large a prediction for the force. Taking into account in the
exact calculations the interplay between the surface geometry and
the optical properties of the materials leads to good agreement
between the experiment and theoretical predictions
\cite{LambrechtPRL2008}, as confirmed later on by another experiment
using a shallow-trenched Si surface and comparing with independent
theoretical calculations \cite{BaoPRL2010}. More recently these
methods have been complemented by another approach based on a
decomposition of the field onto the modes of the structures
\cite{DavidsPRA2010}. See also a recent review on Casimir forces in
structured geometries by Rodriguez \textit{et al}
\cite{RodriguezNaturePh2011}.

The use of nanostructured plates instead of flat surfaces also
allows for the existence of lateral Casimir forces acting tangential
to the surface and of Casimir torques. They arise when the
translational invariance is broken, either due to the anisotropy of
the material \cite{MundayPRA2005} or to surface structures
\cite{RodriguesEPL2006}. Lateral Casimir forces were first
calculated for scalar fields and perfect boundaries imprinted with
one-dimensional (1D) sinusoidal corrugations
\cite{GolestanianPRL1997,EmigPRL2001}, where by 1D we mean
corrugations following a single transverse direction. The case of
electromagnetic fields and non-dissipative metallic surfaces was
studied within the scattering approach in \cite{RodriguesPRL2006}
and later on applied to Casimir-Polder interactions between a
metallic nanostructure and Rb atoms \cite{DalvitPRL2008}. The
Casimir interaction involving single objects of elliptical shapes
have been studied theoretically in
\cite{SerneliusPRA2008,KardarPRA2009,LevinPRL2010} and
experimentally in \cite{DeccaPRA2011}.

Experimental evidence of the lateral Casimir force was given for
sinusoidal corrugations imprinted on a sphere and a plate, with
corrugation periods $d$ much smaller than the surfaces' separation
$L$, and showed its sinusoidal variation with the plates' lateral
displacement \cite{ChenPRL2002}. In this situation good agreement
was found with calculations based on the PFA as the condition $d \ll
L$ was satisfied. More recently, the same group has measured the
lateral force for asymmetric sinusoidal profiles accompanied by
exact calculations \cite{ChiuPRB2009,ChiuPRB2010} using the method
developed in \cite{LambrechtPRL2008}.

An interesting perspective for the lateral Casimir force lies in the
realization of non-contact rack-and-pinion devices or ratchets
\cite{GolestanianPRL2007,EmigPRL2007b,GolestanianPRE2010} and of
non-contact gears \cite{Cavero-PelaezPRD2008,Cavero-PelaezPRD2008b}
which would allow for a force or torque transmission between
corrugated surfaces without bringing them in direct contact with
each other. Typical profiles studied in this context are sinusoidal.

In the present paper we study the Casimir energy between two
nanostructured surfaces as a function of their lateral displacement
from which the magnitude of the lateral force may be easily deduced.
We use the scattering approach
\cite{JaekelJP1991,LambrechtNJP2006,EmigPRL2007} to handle a variety
of different periodic profiles. We present a detailed comparison of
different profiles, such as circular, elliptical, triangular,
trapezoidal and compare the Casimir energies they generate to those
of the commonly used sinusoidal and rectangular profiles. Profiles
with a base angle larger than for the rectangular profile generate
stronger Casimir energies than profiles having a smaller base angle.
While failing in general, we find the PFA to be a good approximation
for the latter profiles displaced by half a corrugation period. Our
results for circular and elliptical profiles can be applied to model
the dispersive interaction between periodic arrays of nanowires or
nanotubes \cite{LinJPCC2008,WangJAC2009,ChenJNN2011}.

\section{Scattering formalism}

For parallel plates, the Casimir force may be written in terms of
scalar reflection coefficients, but for non-planar surfaces the
specular reflection coefficients must be replaced by general
reflection operators that describe the non-specular diffraction by
the surfaces \cite{LambrechtNJP2006}. In the case of two dielectric
lamellar gratings, the Casimir interaction energy per unit area at
zero temperature is given by \cite{LambrechtPRL2008}
\begin{align}
& E=\frac{\hbar}{(2\pi)^3} \iiint
\mathrm{d}^{2}\mathbf{k}_{\perp}\mathrm{d}\xi \ln \det \left(
\mathbf{1}-\mathcal{M}\right)
\label{scatterform} \\
& \mathcal{M} = \mathcal{R}_{1} e^{-\kappa L} \mathcal{R}_{2}
e^{-\kappa L} \notag
\end{align}
The Casimir energy is written in terms of reflection operators
$\mathcal{R}_{1}$ and $\mathcal{R}_{2}$ which describe the
diffraction by the two lamellar gratings. The operator $e^{-\kappa
L}$ accounts for a one-way propagation along the distance $L$
separating the two gratings, with $\kappa = \sqrt{\xi^2 +
\mathbf{k}_{\perp}^2}$ the imaginary longitudinal wave-vector and
$\mathbf{k}_{\perp}$ the transverse wave-vector, where all
quantities are written at imaginary frequencies $\omega =i\xi $
after a Wick rotation. The operator $\mathcal{M}$ thus represents
one round-trip propagation between the two surfaces.

\begin{figure}
\includegraphics[scale=0.37]{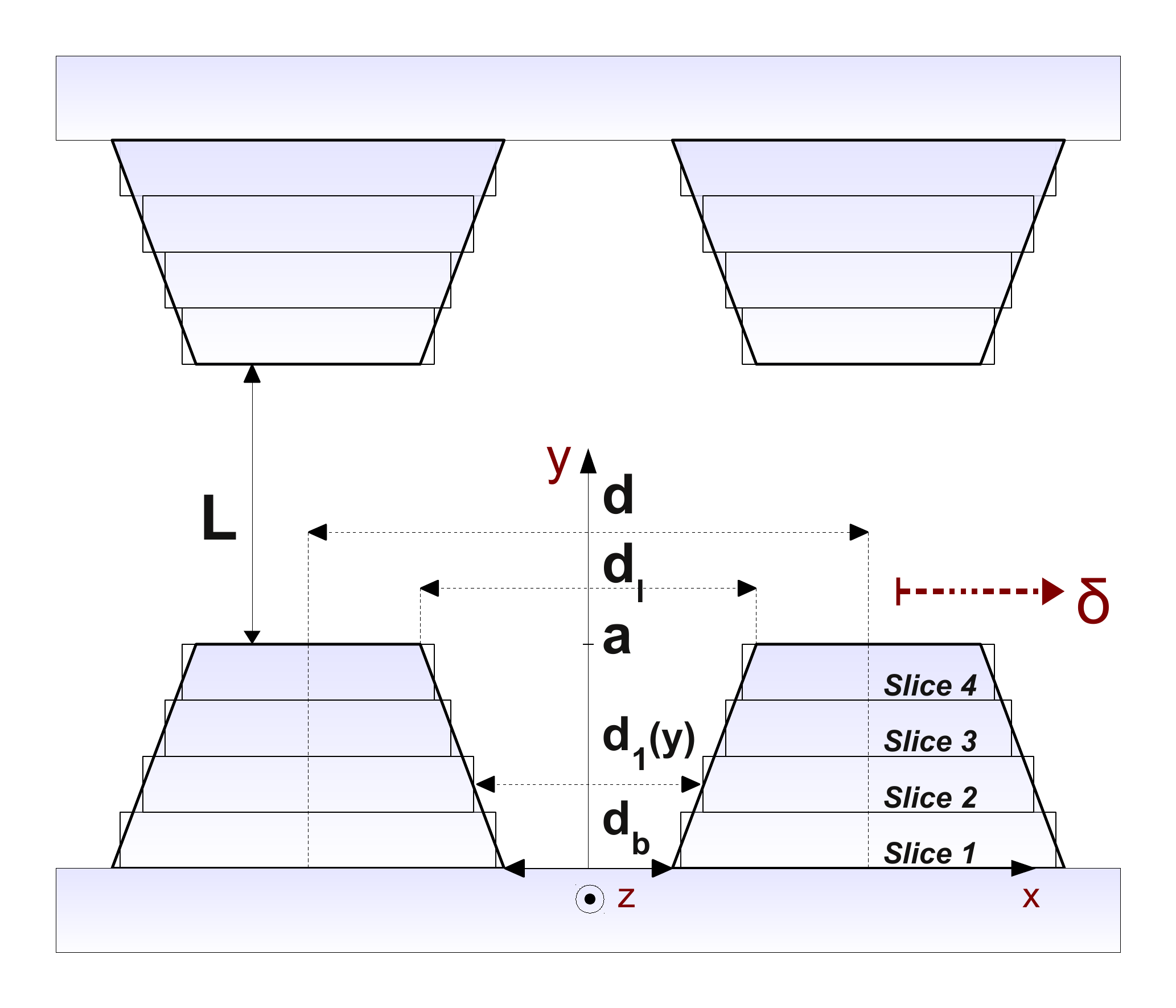}
\caption{\label{Plot1} Two nanostructured surfaces of arbitrary but
symmetric profiles with a period $d$ and corrugation depth $a$ are
divided into $K$ vertical stacks of rectangular slices (here
$K=4$).}
\end{figure}

Here we consider two dielectric lamellar gratings of arbitrary but
symmetric profiles with a period $d$ and corrugation depth $a$
separated by a vacuum slit such as shown in Fig.\ref{Plot1}. In
\cite{LambrechtPRL2008} a formalism has been developed to calculate
the Casimir energy in this configuration for rectangular profiles.
We generalize this approach to arbitrary profiles by dividing each
corrugation line into $K$ horizontal slices vertically stacked on
each other. For a corrugation depth $a$, each slice is treated as a
lamellar rectangular grating whose height along the $y-$axis is
fixed at $a/K$ and whose length along the $x-$axis is given by the
profile's length at the level of the slice.

Let us first recall the main steps of the calculation for
rectangular gratings separated by a vacuum slit. Because of time and
z-invariance, we can write the electric and magnetic fields for
$i=(x,y,z)$ such as :
\begin{eqnarray}
E_{i}(x,y,z,t) = E_{i}(x,y) \exp(ik_{z}z-i \omega t)
\label{eq:e1}
\end{eqnarray}
\begin{eqnarray}
H_{i}(x,y,z,t) = H_{i}(x,y) \exp(ik_{z}z-i \omega t).
\label{eq:e2}
\end{eqnarray}
In the following we will take $\mu_0=\mu=1$, and $c=1$ in vacuum. We
need to find the longitudinal components outside the corrugated
region ($y>a$) and within the transmitted region ($y \leqslant 0$).
In the case of two planar interfaces, we can write the
$z-$components of the fields in the vacuum region as an incident and
reflected field:
\begin{eqnarray}
E_{z}(x,y) = I^{e} e^{ik_x x - ik_y y} + r^e e^{ik_x x + ik_y y}
\label{eq:e3a}
\end{eqnarray}
\begin{eqnarray}
H_{z}(x,y) = I^{h} e^{ik_x x - ik_y y} + r^h e^{ik_x x + ik_y y}
\label{eq:e3b}
\end{eqnarray}
\noindent while in the material region, we can write the
$z-$components of the fields as transmitted fields:
\begin{eqnarray}
E_{z}(x,y) = t^e e^{ik_x x - ik'_y y} \label{eq:e3c}
\end{eqnarray}
\begin{eqnarray}
H_{z}(x,y) = t^h e^{ik_x x - ik'_y y} \label{eq:e3d}
\end{eqnarray}
\noindent where $k_y^2 = \omega^2 - k_x^2 -k_z^2$ and
$k_y'^2=\epsilon \omega^2 -k_x^2 -k_z^2$ are the longitudinal
wave-vectors in vacuum and inside the material. $k_x, k_z$ are the
components of the transverse wave-vector, $\epsilon$ is the
material's frequency dependent permittivity. The coefficients
$I^{e,h}$, $r^{e,h}$, $t^{e,h}$ are respectively the incident
amplitudes and the Fresnel-Stokes reflection and transmission
amplitudes in the $(e,h)$-basis of polarizations, which are defined
by imposing $H_y = 0$ for the $e$-polarization and $E_y = 0$ for the
$h$-polarization.

We may now generalize the above equations for gratings. Outside the
corrugations ($y > a$), the fields are written as a Rayleigh expansion
involving incident and reflected fields of order $p$ and $n$
respectively :
\begin{multline}
E_{z}(x,y)_{y>a} = I_{p}^{e} \exp(i \alpha_{p}x-i \beta_{p}^{(1)}y) \\
+ \sum_{n \in \mathbb{Z}} R_{np}^{e} \exp(i \alpha_{n}x+i \beta_{n}^{(1)}y)
\label{eq:e3}
\end{multline}
\begin{multline}
H_{z}(x,y)_{y>a} = I_{p}^{h} \exp(i \alpha_{p}x-i \beta_{p}^{(1)}y) \\
+ \sum_{n \in \mathbb{Z}} R_{np}^{h} \exp(i \alpha_{n}x+i \beta_{n}^{(1)}y)
\label{eq:e4}
\end{multline}
\noindent For the region inside the material ($y \leqslant 0$), the
transmitted fields are given by :
\begin{eqnarray}
E_{z}(x,y)_{y \leqslant 0} = \sum_{n \in \mathbb{Z}} T_{np}^{e} \exp(i \alpha_{n}x-i \beta_{n}^{(2)}y)
\label{eq:e5}
\end{eqnarray}
\begin{eqnarray}
H_{z}(x,y)_{y \leqslant 0} = \sum_{n \in \mathbb{Z}} T_{np}^{h}
\exp(i \alpha_{n}x-i \beta_{n}^{(2)}y), \label{eq:e6}
\end{eqnarray}
where we have used:
\begin{eqnarray}
\alpha_{p} & = & k_{x}+2\pi p/d \\
\alpha_{n} & = & k_{x}+2\pi n/d \\
\beta_{p}^{(1)2} & = & \omega^{2}-k_{z}^2-\alpha_{p}^2 \\
\beta_{n}^{(1)2} & = & \omega^{2}-k_{z}^2-\alpha_{n}^2 \\
\beta_{n}^{(2)2} & = & \epsilon \omega^{2}-k_{z}^2-\alpha_{n}^2.
\label{eq:e7}
\end{eqnarray}
$I_{p}$, $R_{np}$, and $T_{np}$ are now the incidence, reflection,
and transmission matrix elements respectively. $n=0$ corresponds to
a specular reflection. By symmetry, the other field components of the
electric and magnetic fields can each be expressed through the
z-components of both fields, following Maxwell's equations.

We now need to determine the reflection coefficients $R_{np}$ of the
rectangular corrugated gratings. To this aim we first rewrite
Maxwell's equations inside the corrugated region $0<y \le a$ through
the set of first-order differential equations $\partial_{y}
\mathbf{F}=\mathbf{M} \mathbf{F}$, for
$\mathbf{F}^{\top}=(E_{x},E_{z},H_{x},H_{z})$ and $\mathbf{M}$ a
constant square matrix of dimension $8N+4$. The solution for the
fields is then of the form :
\begin{eqnarray}
\mathbf{F}(y)=e^{\mathbf{M}y} \mathbf{F}(0)
\label{eq:e8bb}
\end{eqnarray}
with :
\begin{eqnarray}
\mathbf{M} =
\left( \begin{array}{cccc}
0                & 0              &
\frac{-ik_z \alpha_{n}}{\epsilon\omega} &
-i\frac{\epsilon\omega^2-\alpha_n^2}{\epsilon\omega} \\
0                & 0              &
i \frac{\epsilon\omega^2 - k_z^2}{\epsilon\omega} &
\frac{ik_z \alpha_n}{\epsilon\omega} \\
\frac{ik_z \alpha_n }{\omega} &
i\frac{\epsilon\omega^2 - \alpha_n^2}{\omega} & 0              & 0              \\
- i \frac{\epsilon\omega^2 - k_z^2}{\omega} &
\frac{-i k_z \alpha_{n}}{\omega} & 0              & 0              \\
\end{array} \right)
\label{eq:e8c}
\end{eqnarray}
where the elements appearing in matrix $\mathbf{M}$ are block
matrices of dimension $2N+1$. We can write the fields inside the
corrugation region and match them through continuity relations for
each $E_{x}$, $E_{z}$, $H_{x}$, $H_{z}$, with equation
(\ref{eq:e8bb}), at boundary $y=a$ for $y>a$, and at boundary $y=0$
for $y \leqslant 0$. This allows to find the vectors $\mathbf{F}(a)$
and $\mathbf{F}(0)$, which can be written as a product of a matrix
and the vector of variables $X$ :
\begin{eqnarray}
\mathbf{F}(a) = TX+Y \hspace{5mm} \text{and} \hspace{5mm} \mathbf{F}(0) = SX
\label{eq:e8d}
\end{eqnarray}
with $X^{T}=(R_{np}^e,R_{np}^h,T_{np}^e,T_{np}^h,\ldots)$, and $Y$
being the variable-independent term including the polarization of
the incident waves, since we must take into account the two
polarizations, $e$ and $h$ independently: $I_p^{(e)}=1$ and
$I_p^{(h)}=0$ for electric waves ($H_z=0$), and $I_p^{(e)}=0$ and
$I_p^{(h)}=1$ for magnetic waves ($E_z=0$). $Y$ hence characterizes
the two separate solutions for $e$- and $h$-waves. The solution is
then of the form:
\begin{eqnarray}
X=(e^{\mathbf{M}a}S-T)Y.
\label{eq:e8e}
\end{eqnarray}
\noindent We have :
\begin{eqnarray}
X \left( I_p^{(e)}=1, I_p^{(h)}=0 \right) =
\left( \begin{array}{c}
R_{np}^{(e,e)} \\
R_{np}^{(h,e)}  \\
\vdots \\
\end{array} \right)
\label{eq:e8f}
\end{eqnarray}
\begin{eqnarray}
X \left( I_p^{(e)}=0, I_p^{(h)}=1 \right) =
\left( \begin{array}{c}
R_{np}^{(e,h)} \\
R_{np}^{(h,h)}  \\
\vdots \\
\end{array} \right)
\label{eq:e8g}
\end{eqnarray}
\noindent so that we obtain the reflection matrix for each grating:
\begin{eqnarray}
R (\omega) =
\left( \begin{array}{cccc}
R_{np}^{(e,e)}     &
R_{np}^{(e,h)}     \\
R_{np}^{(h,e)}     &
R_{np}^{(h,h)}     \\
\end{array} \right)
\label{eq:e8h}
\end{eqnarray}

After making use of Cauchy's argument principle and normalizing the
frequency by $c$, we arrive to the exact expression of the Casimir
energy between the two gratings on a unit cell of period $d$ and
unit length in the $y$-direction, with $R_{1}(i\xi)$ and
$R_{2}(i\xi)$:
\begin{multline}
E = \frac{\hbar d c}{8\pi^{3}} \int_{\mathbb{R}^{+ \ast}} d\xi
\int_{\mathbb{R}} dk_{z}
\int_{-\pi/d}^{\pi/d} dk_{x} \\
\\
\times \ln \det[\mathbf{1}-R_{1}(i\xi)
e^{-\mathcal{K}L}R_2(i\xi)e^{-\mathcal{K}L}] \label{eq:e9}
\end{multline}
with $\mathcal{K} = \mathrm{diag}(\sqrt{\xi^2 + k_y^2 +
[k_x+(2m\pi/d)]^2})$ and $m=-N,\ldots,+N$. For the sake of clarity
we have explicitly re-introduced the speed of light $c$ here.

\section{Arbitrary profiles}

We will now consider gratings of arbitrary symmetric profiles. The
difference with the rectangular gratings appears in the parameter
$d_{1}$, which will now depend on $y$. Arbitrary profiles defined by
$d_{1}(y)$ can be divided into $K$ slices, each of rectangular
shape, as described in Fig. \ref{Plot1}. For each slice $(i)$, the
spacing between the corrugation ridges is $d_{1}^{(i)}$ and the
former scattering formalism for rectangular corrugations can be
applied. More specifically, a differential equation $\partial_{y}
\mathbf{F}=\mathbf{M}^{(i)} \mathbf{F}$ akin to equation
(\ref{eq:e8bb}) can be solved within each slice $(i)$ to relate the
fields at boundary $y=i \frac{a}{K}$ and $y=(i+1) \frac{a}{K}$.

In a similar way than for the case $K=1$ above, the field at $y=a$ is thus related to the field at $y=0$ via the relation :
\begin{eqnarray}
\mathbf{F}(a)=\left[ \prod_{i=K}^{1} e^{\mathbf{M}^{(i)} \frac{a}{K}} \right] \mathbf{F}(0)
\label{eq:e10}
\end{eqnarray}
where the product $\prod$ runs from $i=K$ to $i=1$.

\begin{figure}
\includegraphics[scale=0.4]{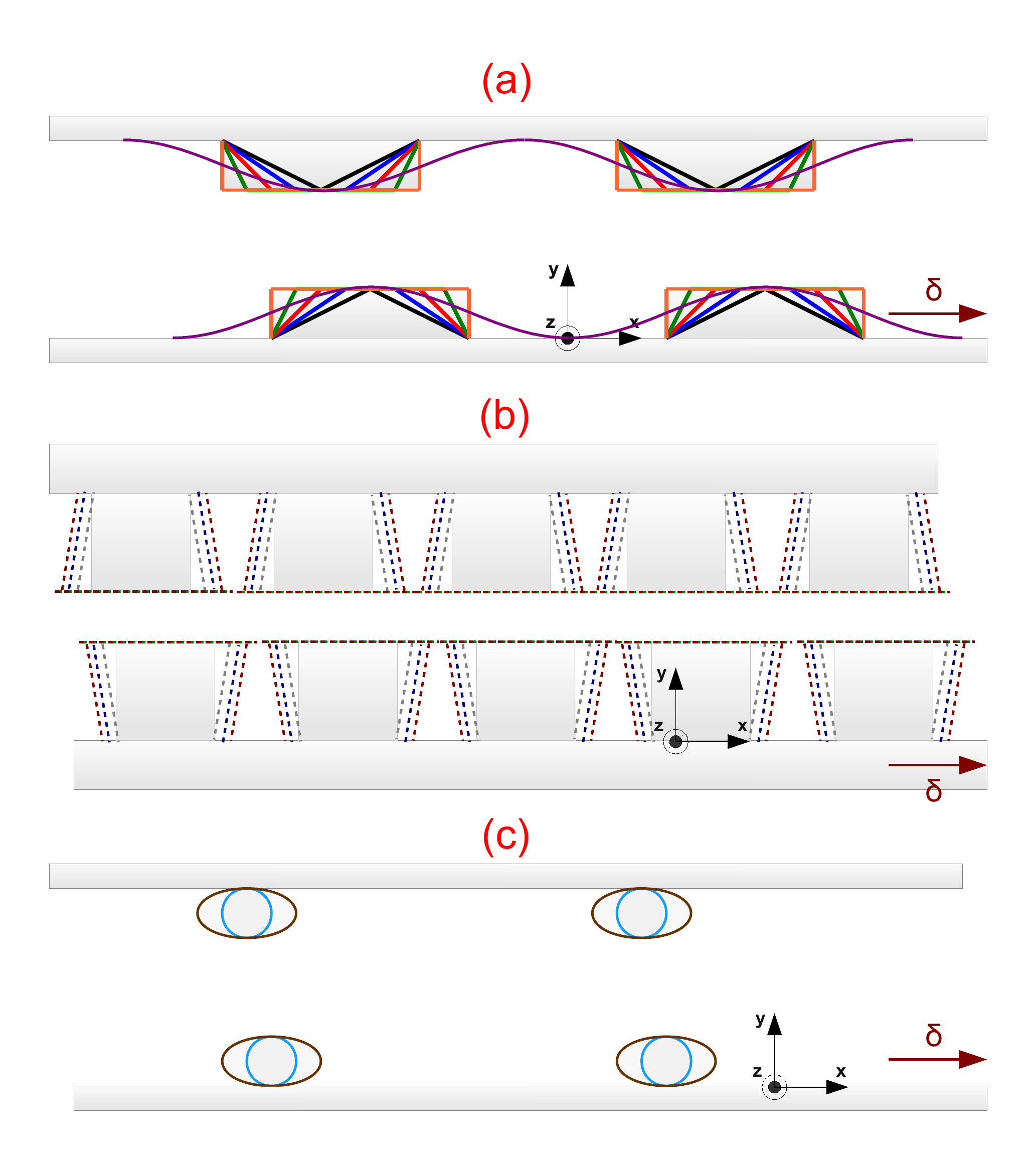}
\caption{\label{Plot2} Two-dimensional perspective on the different
considered periodic gratings. (a) and (c) are at the same scale,
whereas the scale of (b) has been increased by a factor two.}
\end{figure}

\begin{figure}
\includegraphics[scale=0.3]{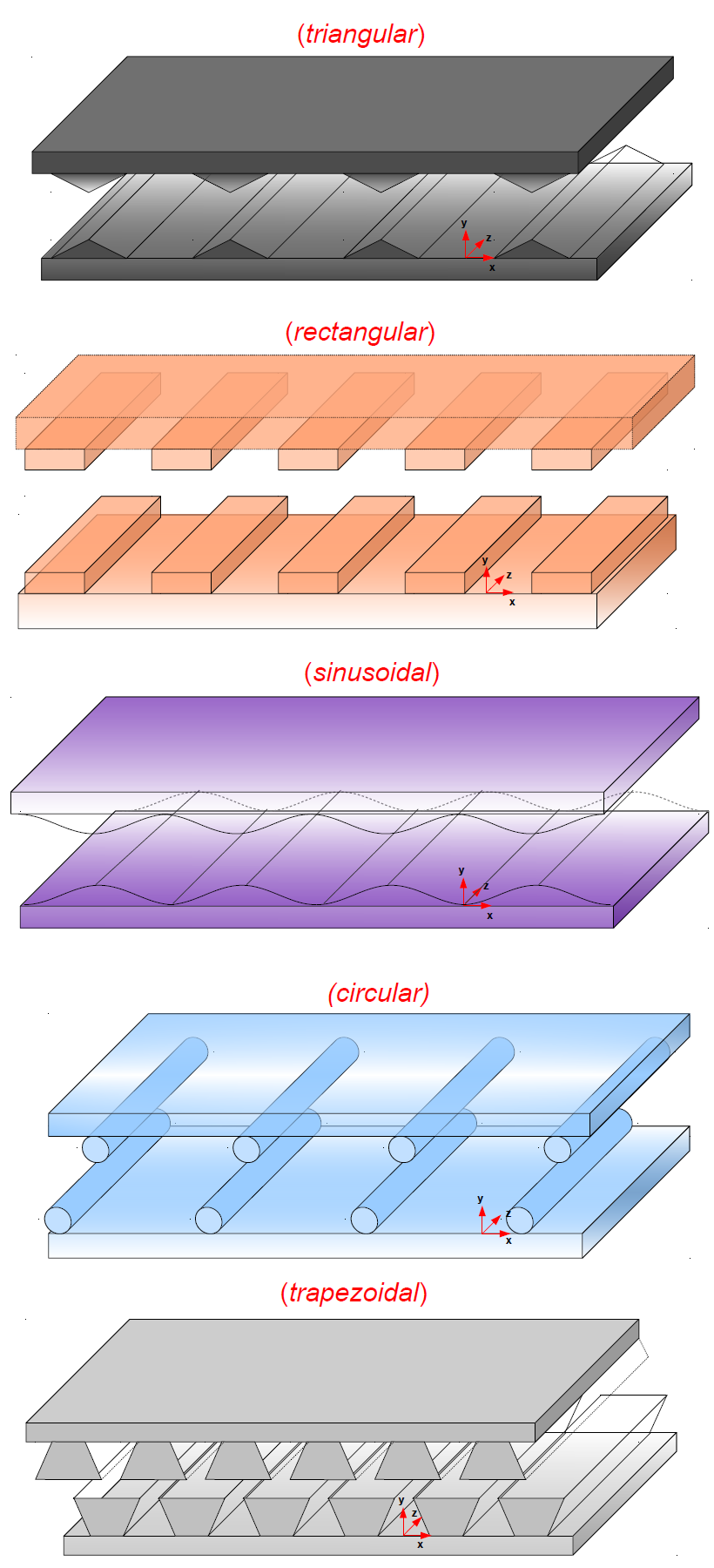}
\caption{\label{Plot3} Three-dimensional perspective on the
triangular, rectangular, sinusoidal, circular, and trapezoidal periodic
profiles shown in Fig. \ref{Plot2}. The profiles are here represented at
the same scale.}
\end{figure}

Hence a correct parametrization of the quantity $d_{1}$ as a
function of $y$ allows one to generate arbitrary symmetric profiles
for the corrugations. The profiles that we will study as examples in
the following are shown on Fig. \ref{Plot2} as cross sections while
on Fig. \ref{Plot3} the grating structure becomes more apparent.
Triangular profiles with $d_l=d$ as seen on Fig. \ref{Plot2}a and
Fig. \ref{Plot3} are generated by the function $d_{1}(y)=\frac{d}{a}
y$, sinusoidal profiles by $d_1(y)= (d/\pi) \arccos[1-(2y/a)]$. We
also study two types of trapezoidal profiles, having a base angle
smaller than 90$^\circ$ (Fig. \ref{Plot2}a) on the one hand, and
having a base angle larger than 90$^\circ$ (Fig. \ref{Plot2}b) on
the other hand. They are characterized by $d_{b}<d_{l}$ and
$d_{b}>d_{l}$ respectively and are both generated by the function
$d_{1}(y)=(d_{l}-d_{b})y/a + d_{b}$. Ellipsoid profiles along the
$x$- or $y$-axis are generated by $d_{1}(y)=d - \frac{2R}{r}
\sqrt{r^{2}-(y-Y)^{2}}$ (Fig. \ref{Plot2}c) and $d_{1}(y)=d -
\frac{2r}{R} \sqrt{R^{2}-(y-Y)^{2}}$ respectively, for $R$ and $r$
being the major and minor axes of the ellipse, and $Y$ being the
value of the $y$-coordinate of the ellipse center. Circular periodic
profiles are also generated by these expressions (as seen on Fig.
\ref{Plot2}c and Fig. \ref{Plot3}), with $R=r$ being the radius in
the $xy$-plane. Note that what we call for example a circular or
elliptical periodic profile is in fact a geometry where equally
spaced parallel wires of circular or elliptical cross-section cover
the surfaces, as shown on Fig. \ref{Plot3}.

\begin{figure}
\includegraphics[scale=0.77]{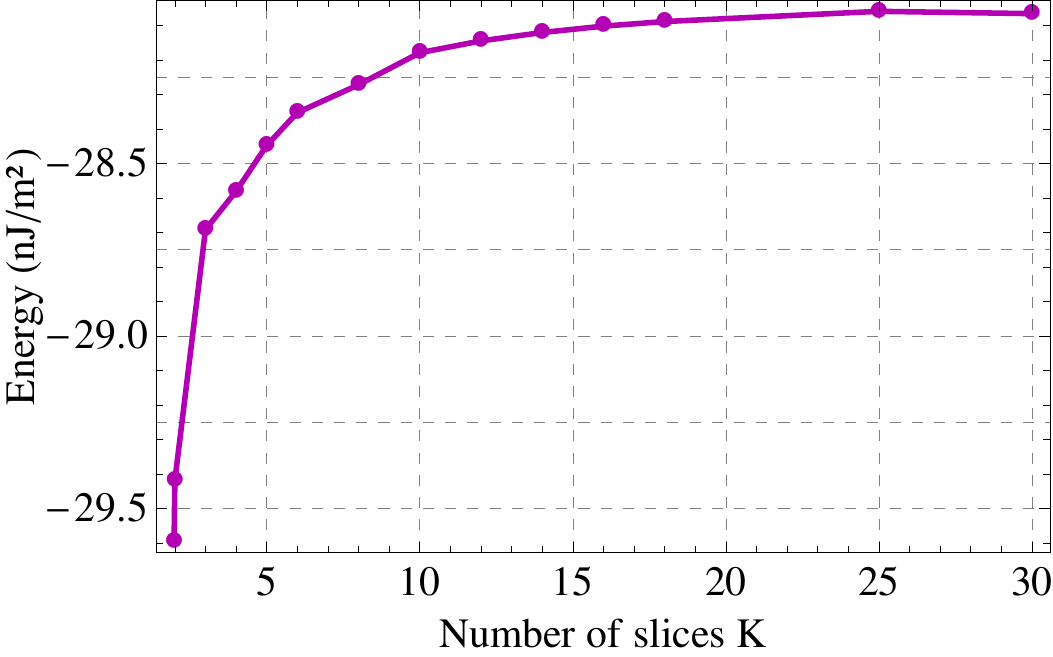}
\caption{\label{Plot4} Dependence of the Casimir energy on the
number of slices $K$ for two triangular gratings, with $L=100$ nm,
$d=400$ nm, $a=50$ nm, and $d_{1}(y)=4y+200$.}
\end{figure}

Obviously a given profile will be better fitted for greater numbers
of slices $K$. The number of slices thus determines the accuracy of
the overall model. Fig. \ref{Plot4} shows the Casimir energy as a
function of the number of slices for two triangular profiles at a
distance $L=100$ nm, for a grating period $d=400$ nm, corrugation
depth $a=50$ nm, and distance between the ridges $d_1(y)=4y+200$.
For this example, convergence sets in for $K \approx 20$.

In the following we will also compare the results of the scattering
theory presented here with the proximity force
approximation~\cite{DerjaguinKollZ1934}. The PFA comes from the
weighted sum of the planar normal contributions $E_{\text{PP}}(L)$
depending on the local distances $L$ within each period, and hence
also on the lateral displacement $\delta$ between the gratings. If
we express the shapes of the arbitrary periodic gratings in an
analytical form such as $y=f(x,\delta)$ for the lower grating and
$y=L+2a -f(x,\delta=0)$ for the upper grating in the $xy$-plane
shown on Fig. \ref{Plot1}, we can then define the function
$h(x,\delta)=L+2a -f(x,\delta=0)-f(x,\delta)$ in order to express
the local distance of separation between the two profiles. Dividing
the period $d$ in a number $N \rightarrow \infty$ of intervals of
individual widths $d/N \rightarrow 0$, we then obtain a general
expression of the Casimir energy in the PFA for arbitrary gratings
as a function of lateral displacement $\delta$ :
\begin{eqnarray}
E^{\text{PFA}}(L,\delta) & = & \frac{1}{d} \int_{0}^{d}  E_{\text{PP}} \left( h(x,\delta) \right) dx \\
& = & \frac{1}{N} \sum_{i=1}^{N} E_{\text{PP}} \left( L= h
\left(x=i\frac{d}{N},\delta \right) \right) \notag \label{eq:e10b}
\end{eqnarray}

\section{Casimir energy for arbitrary periodic gratings}

We now evaluate numerically the Casimir energy for several types of
profiles as a function of the surfaces' relative lateral
displacement $\delta$. The material chosen for these profiles is
intrinsic silicon, which can be described by a Drude-Lorentz
function \cite{BergstromACIS1997}
\begin{eqnarray}
\epsilon(i\xi) = \epsilon_{\infty}+
\frac{(\epsilon_{0}-\epsilon_{\infty}) \omega_0^2}{\omega^2 +
\omega_0^2}\label{epsilonapprox}
\end{eqnarray}
The numerical values are determined by realizing that at
low-frequencies the dielectric function of intrinsic silicon
approaches the constant value $\epsilon_0=11.87$ while with
increasing frequency it is nearly constant and falls off only for
high frequencies above a cut-off frequency $\omega_0 \approx 6.6
\cdot 10^{15}$ rad/s. For high frequencies it reaches the asymptotic
value $\epsilon_{\infty}=1.035$ \cite{PirozhenkoPRA2008}.

\begin{figure}
\includegraphics[scale=0.55]{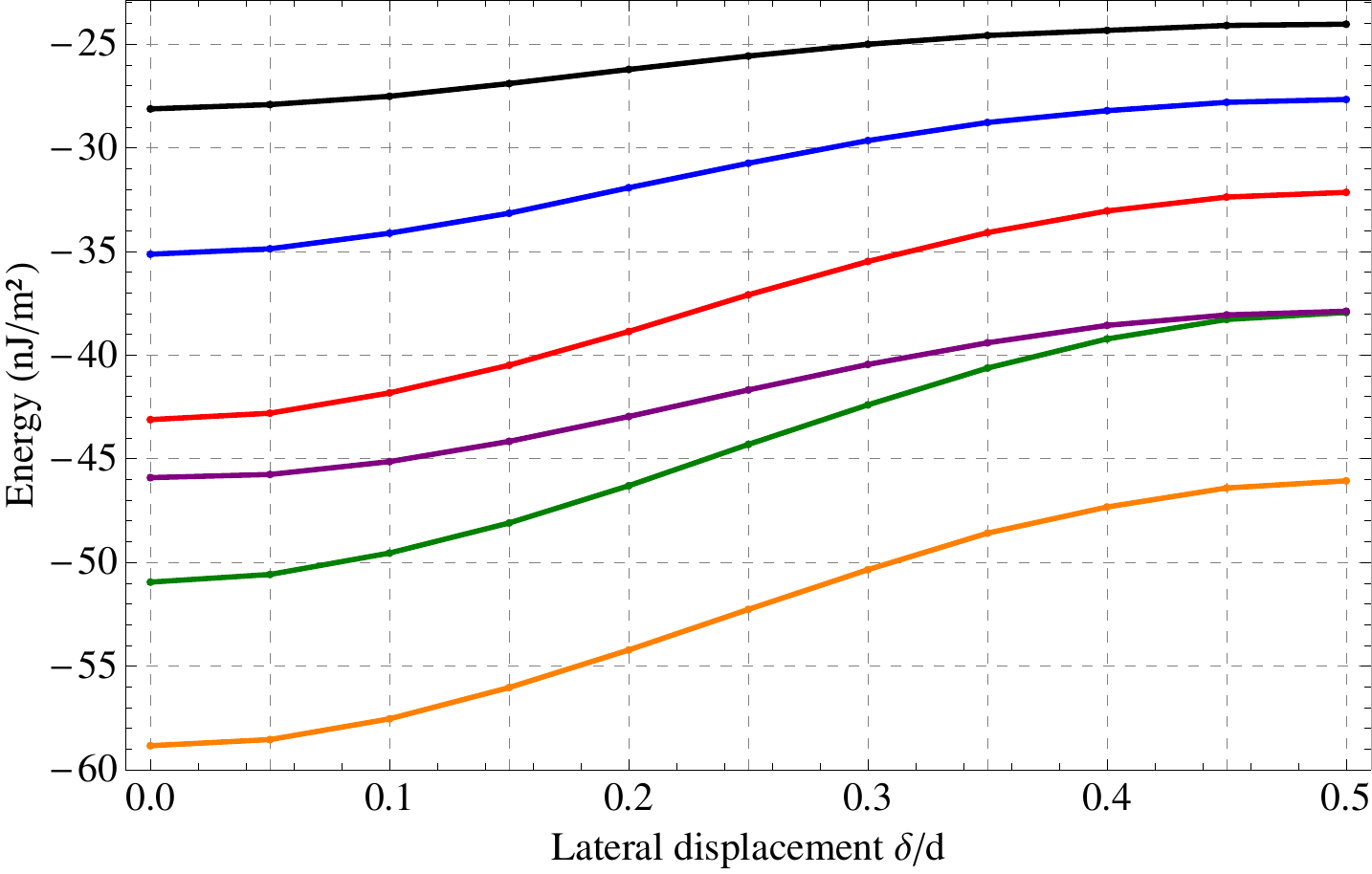}
\caption{\label{Plot5} Dependence of the Casimir energy on the
lateral displacement between two corrugated periodic profiles of
intrinsic silicon for $L=100$ nm, $K = 20$, $d=400$ nm, $a=50$ nm,
and from top to bottom, $d_{1}=4y+200$ (triangular in black),
$d_{1}=3y+200$ (blue), $d_{1}=2y+200$ (red), sinusoidal profile
(purple), $d_{1}=y+200$ (green) and $d_{1}=200$ (rectangular in
orange). The sinusoidal profile is plotted for the same parameters,
except $d_1= (d/\pi) \arccos[1-(2y/a)]$.}
\end{figure}

We first study the transition from triangular over trapezoidal to
rectangular profiles, such as sketched by the colored shapes of Fig.
\ref{Plot2}a. Those profiles are parametrized from top to bottom by
$d_{1}(y)=4y+200$ (triangular: black), $d_{1}(y)=3y+200$
(trapezoidal: blue), $d_{1}(y)=2y+200$ (trapezoidal: red),
$d_{1}(y)=y+200$ (trapezoidal: green), and $d_{1}(y)=200$
(rectangular: orange) respectively. For these profiles the top
distance $d_l$ decreases successively from $d_l = d = 400$ nm to
$d_l = d_b = 200$ nm by steps of $50$ nm. The different Casimir
energies for $L=100$ nm, $d=400$ nm, $a=50$ nm, and a number of
slices $K=20$, are depicted in Fig. \ref{Plot5}.
Clearly the Casimir energy increases as $d_{l}$ decreases, and this
is especially true at $\delta=d/2$, meaning that both the Casimir
energy and its modulation over lateral displacement are larger for
smaller $d_{l}$, i.e. for rectangular gratings and smaller for
larger $d_l$ such as triangular profiles. Note also the behavior of
the sinusoidal profile at $\delta=d/2$, which shows that such
profiles are much less sensitive to lateral displacement than
triangular profiles or rectangular gratings.

\begin{figure}
\includegraphics[scale=0.55]{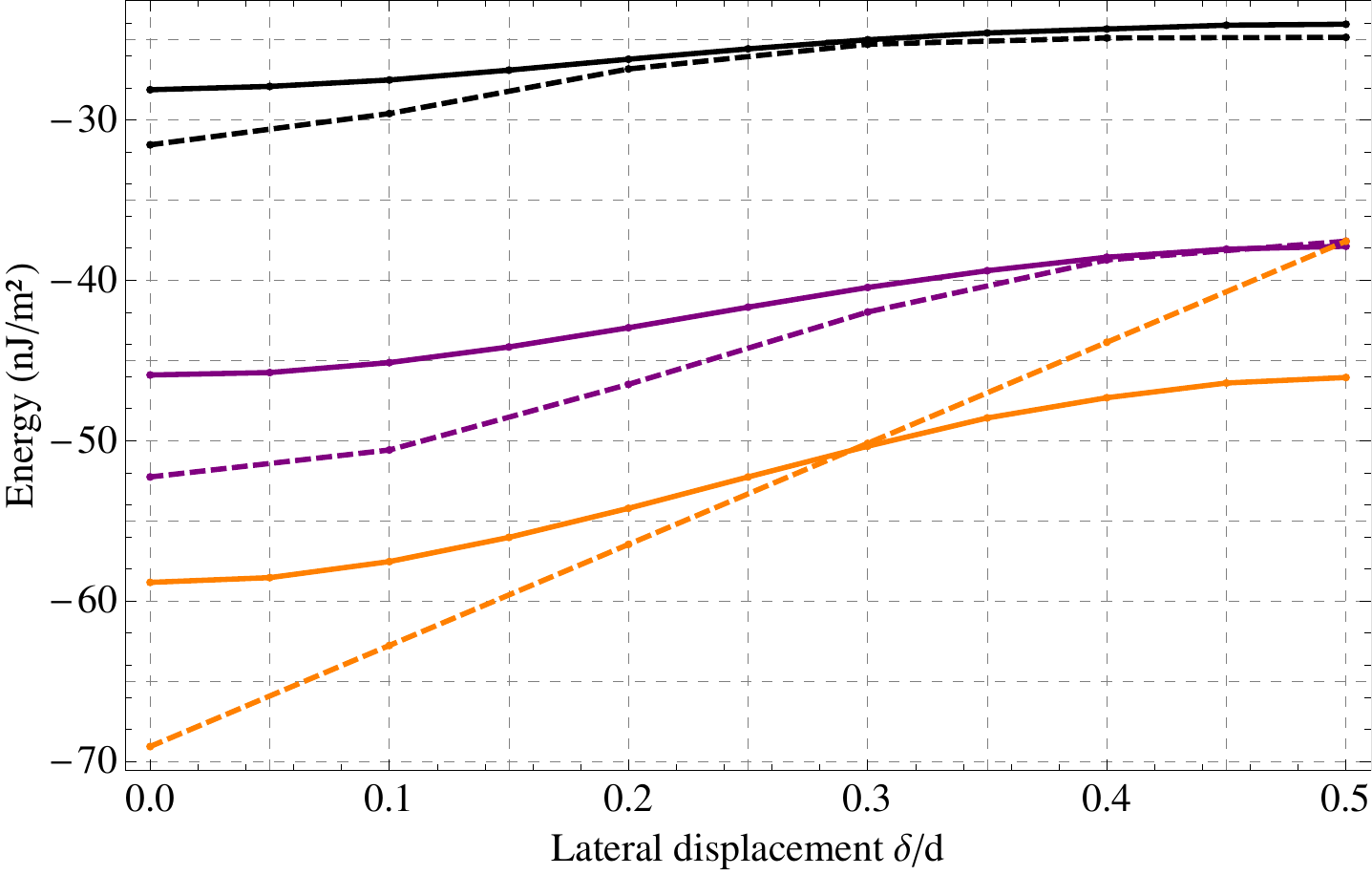}
\caption{\label{Plot6} Comparison between the exact results from
scattering theory (solid lines) and PFA predictions (dashed lines)
for the triangular (black), sinusoidal (purple), and rectangular
(orange) profiles of Fig. \ref{Plot5}.}
\end{figure}

On Fig. \ref{Plot6} 
we show a comparison between the exact results from the scattering
formalism (solid lines) and the predictions from PFA (dashed lines)
for the triangular, sinusoidal, and rectangular profiles (from top
to bottom). Regardless of the considered profile, the PFA fails to
correctly describe the situation of no lateral displacement
$\delta=0$. The error ratio
$E_{\text{PFA}}-E_{\text{scattering}}/E_{\text{PFA}}$ there is
approximately equal to $11\%$ for triangular, $12\%$ for sinusoidal,
and $15\%$ for rectangular gratings. As one shifts $\delta$ to
half-a-period $\delta =d/2$, the relative error for rectangular
gratings passes through zero to reach basically the same value of
opposite sign at $\delta =d/2$. For rectangular gratings PFA thus
underestimates the Casimir energy when corrugation maxima face
maxima and overestimates it when maxima face minima. However, PFA
turns out to give valid predictions for triangular and sinusoidal
gratings when they have a relative lateral shift of half a period.
This is due to the fact that, unlike for rectangular gratings, in
this situation the two triangular or sinusoidal profiles become
parallel to each other.

\begin{figure}
\includegraphics[scale=0.62]{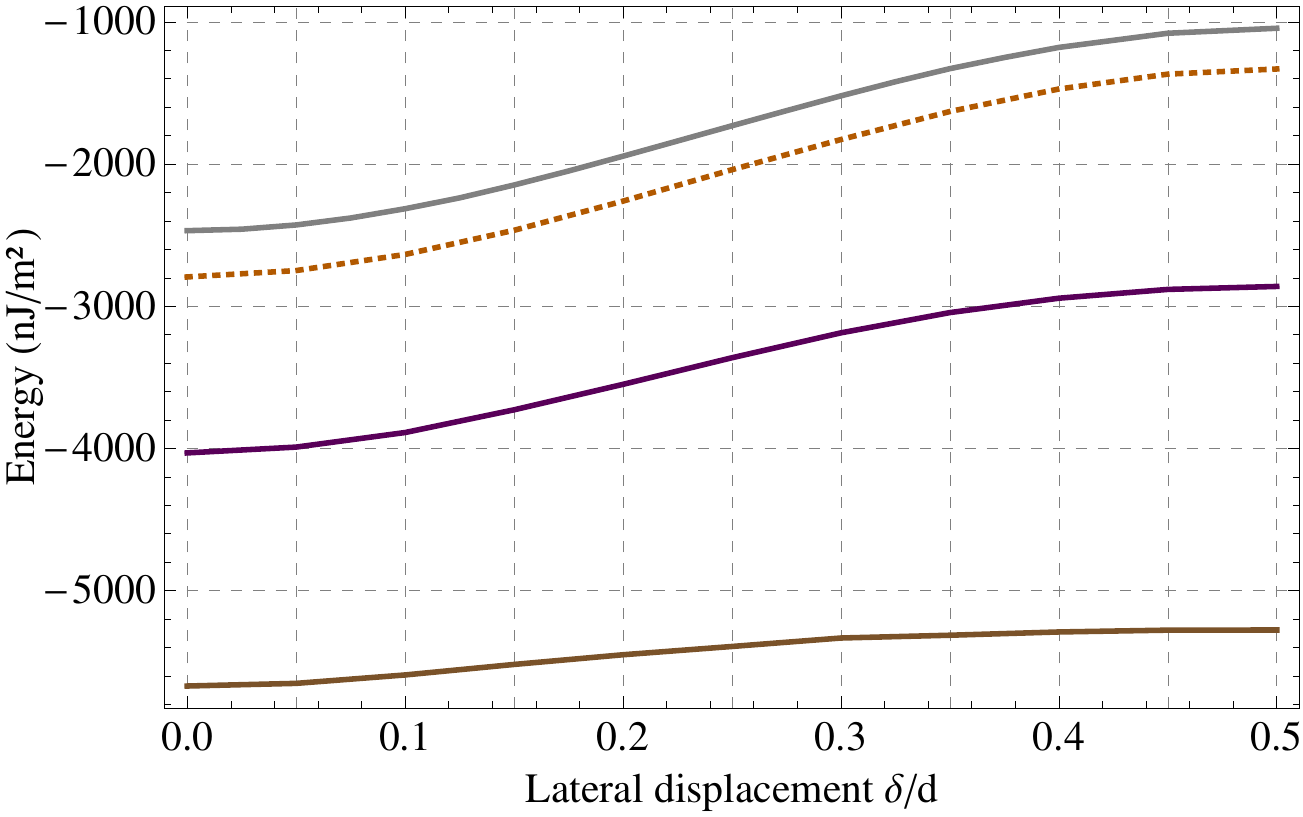}
\caption{\label{Plot7} Dependence of the Casimir energy on the
relative lateral displacement of two trapezoidal periodic profiles
of intrinsic silicon for $L=20$ nm, $K = 20$, $d=100$ nm, $a=50$ nm,
and, from top to bottom, $d_{1}=-0.5y+85$ (grey), $d_{1}=-0.5y+65$
(deep purple), $d_{1}=-0.5y+45$ (brown). This is compared to
rectangular profiles with $d_{1}=d_{l}=60$ nm (orange dotted line).
While we varied the spacing between the corrugations $d_1$, the
grating period $d$ was kept constant.}
\end{figure}

We now study the particular case of trapezoidal profiles for which
$d_{b}>d_{l}$, with a base angle larger than rectangular gratings.
We illustrate the results on Fig. \ref{Plot7} for $L=20$ nm, $d=100$
nm, $a=50$ nm, and a number of slices $K=20$. The profiles are
enlarged on the top while keeping the base constant via a
parametrization $d_{1}=-0.5y+85$ (grey), $d_{1}=-0.5y+65$ (deep
purple), $d_{1}=-0.5y+45$ (brown) from top to bottom. The dotted
orange line corresponds to the Casimir energy between rectangular
gratings with $d_{1}=d_{l}=60$ nm so that the corrugations are as
wide as the top of the trapezoidal profile $d_{1}=85-y/2$. The
difference in the Casimir energies associated with the rectangular
profiles and the trapezoidal profiles, which have the same surface
exposed in the near field, implies the existence of highly
non-trivial mode contributions in the vicinity of the bases of the
trapezoidal gratings.

\begin{figure}
\includegraphics[scale=0.64]{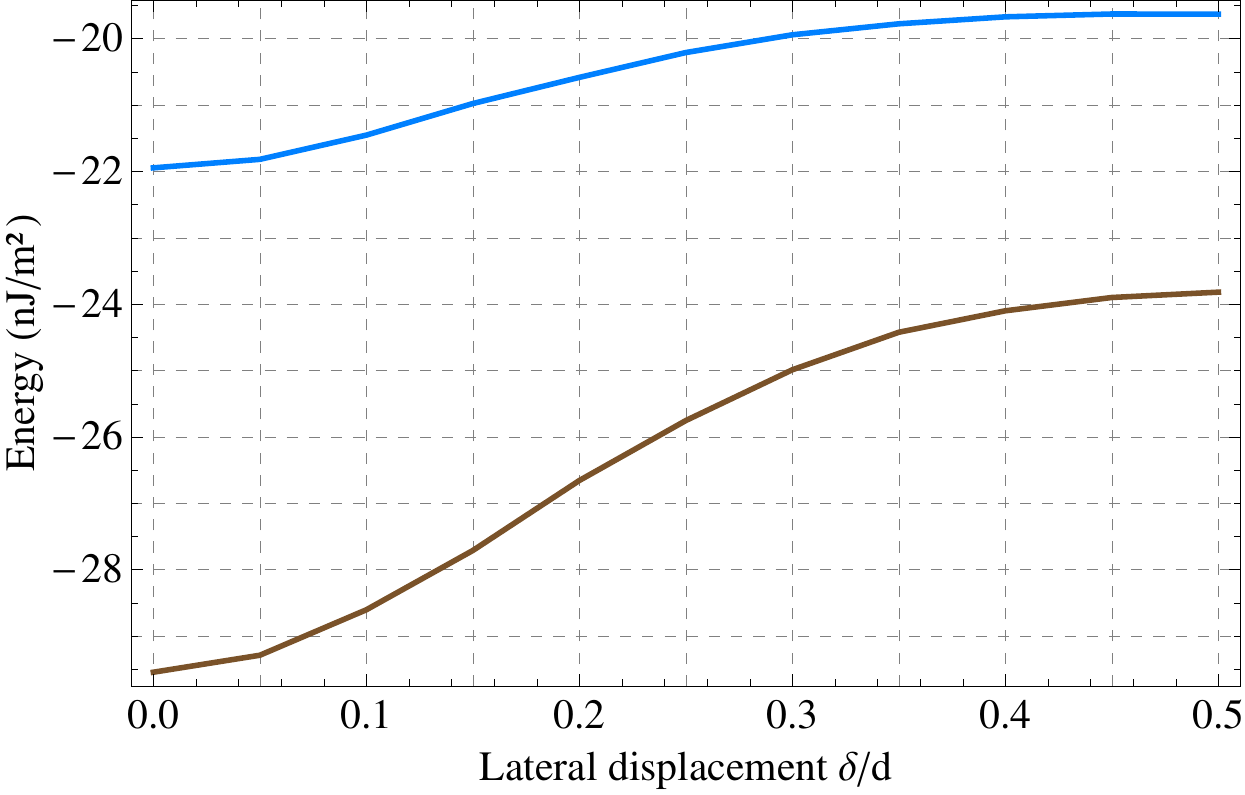}
\caption{\label{Plot8} Dependence of the Casimir energy on the
relative lateral displacement between two periodic profiles shaped
as circular (light blue) and ellipsoid (brown) from top to bottom
for $L=100$ nm, $K=15$, $d=400$ nm, $a=50$ nm, and $d_{1}=400-2
\sqrt{50y - y^{2}}$ (circular with $R=r=Y=25$ nm), and $d_{1}=400-4
\sqrt{50y - y^{2}}$ (ellipsoid with $r=Y=25$ nm and $R=50$ nm). }
\end{figure}

Finally we consider the Casimir interaction between two elliptical
and two circular profiles, as shown in Fig. \ref{Plot8}, for $L=100$
nm, $d=400$ nm, $a=50$ nm, and a number of slices $K=15$.
The elliptical profile (lower curve in brown) has a major axis
$R=50$ nm along $x$, a minor axis $r=25$ nm, and its origin at
$y=Y=25$ nm, so that $d_{1}(y)=400-4 \sqrt{50y - y^{2}}$. The
circular profile (upper curve in blue) has a radius $R=25$ nm and
its center at $y=Y=25$ nm, such that $d_{1}=400-2 \sqrt{50y -
y^{2}}$. The Casimir energy increases with the ratio $R/r$. This is
again especially true at $\delta=d/2$. We may also note that
compared to the profiles studied in Fig. \ref{Plot5}, the energy
varies much more rapidly over $\delta$ for the ellipsoid and
circular profiles. This could be a consequence of the concave nature
of these shapes for $y<Y$ and could potentially affect the lateral
Casimir force applications mentioned earlier
\cite{GolestanianPRL2007,EmigPRL2007b,GolestanianPRE2010,Cavero-PelaezPRD2008,Cavero-PelaezPRD2008b}.
At fixed geometrical parameters and distances, the Casimir energy is
overall much smaller for these profiles than for the triangular and
trapezoidal shapes discussed before.

\section{Conclusion}

We have studied the dependence of the Casimir energy on the lateral
displacement for different arbitrary periodic gratings, ranging from
triangular and sinusoidal profiles, to trapezoidal, circular and
ellipsoid shapes. Concerning the trapezoidal profiles, we find that
at the same distance $L$, grating period $d$, and corrugation depth
$a$, the Casimir energy and its lateral modulation increase from
triangular profiles over rectangular ones to those whose base angle
is larger than 90$^{\circ}$. This seems a consequence of the fact
that for a given surface, the Casimir energy increases when the
exposed surfaces in near-field is increased. Profiles with large
base angles or at least rectangular profiles seem therefore more
promising for lateral Casimir force, Casimir torque or other
non-contact devices
\cite{GolestanianPRL2007,EmigPRL2007b,GolestanianPRE2010,Cavero-PelaezPRD2008,Cavero-PelaezPRD2008b}
than sinusoidal or triangular profiles. However, the exposed surface
in near-field is not sufficient to estimate reliably the magnitude
of the Casimir interaction even for small corrugation depth ($a <
L$), as shown by the comparison of the Casimir energies between
rectangular gratings and large base trapezoidal profiles. While
failing to describe the Casimir interaction correctly in general,
interestingly PFA gives valid predictions for sinusoidal and
triangular gratings when they are relatively displaced by half a
grating period. An interesting topic to further investigate would be
to parametrize the profiles such that the lateral displacement
$\delta$ also depends on $y$, thus generating asymmetric profiles
\cite{ChiuPRB2010}.

We thank the European Science Foundation (ESF) within the activity
\textit{New Trends and Applications of the Casimir Effect}
(www.casimir-network.com) for support.

\bibliographystyle{apsrev4-1}
\bibliography{biblioCorrArb}

\end{document}